\documentclass[11pt,dvips]{article}
\usepackage{epsfig,times} 
\usepackage{picinpar}
\usepackage{wrapfig}
\usepackage{spez}
\usepackage{floatflt}
\makeatletter
\renewcommand{\section}{\@startsection{section}{1}{0in}
	{0.4\baselineskip}{0.1\baselineskip}{\Large\bf}}
\renewcommand{\subsection}{\@startsection{subsection}{2}{0in}
	{0.25\baselineskip}{-\baselineskip}{\large\bf}}
\renewcommand{\subsubsection}{\@startsection{subsubsection}{3}{0in}
	{0.1\baselineskip}{-\baselineskip}{\normalsize\bf}}
\makeatother
\newcommand{\tcorr}{t_{\rm corr}}
\newcommand{\pperp}{p_{\scriptscriptstyle \bot}}
\newcommand{\eqb}{\begin{eqnarray}}
\newcommand{\eqe}{\end{eqnarray}}

\begin{document}
\thispagestyle{myheadings}
\markright{OG 3.3.22}
\begin{center}
{\LARGE \bf Monte-Carlo simulation of particle acceleration \\ 
in braided magnetic fields}
\end{center}
\begin{center}
{\bf U.D.J. Gieseler$^{1}$, and J.G. Kirk$^{2}$}\\
{\it $^{1}$University of Minnesota, Department of Astronomy,
        116 Church St. S.E., Minneapolis, MN 55455, U.S.A.\\
$^{2}$Max-Planck-Institut f\"ur Kernphysik,
Postfach 10 39 80, 69029 Heidelberg, Germany}
\end{center}
\begin{center}
{\large \bf Abstract\\}
\end{center}
\vspace{-0.5ex}
Supernova remnants are expected to contain braided (or stochastic) magnetic 
fields, which are in some regions directed mainly perpendicular to the shock 
normal. For particle acceleration due to repeated shock crossings, the 
transport in the direction of the shock normal is crucial. The mean squared 
deviation along the shock normal is then proportional to the square root of 
the time. This kind of anomalous transport is called {\em sub}-diffusion.
We use a Monte-Carlo method to examine this non-Markovian transport and the 
acceleration. As a result of this simulation we are able to examine the 
propagator, density and pitch-angle distribution of accelerated particles,
and especially the spectral properties. 
These are in broad agreement with analytic predictions for both
the {\em sub}-diffusive and the diffusive regimes, but the steepening of the
spectrum predicted when changing from diffusive to {\em sub}-diffusive 
transport is found to be even more pronounced than predicted.
\vspace{1ex}

\section{Introduction}
\label{intro.sec}
The acceleration of high energy particles in astrophysical plasmas
is a transport process in configuration and momentum space. In describing
the acceleration of charged particles in a magnetised plasma, most analytical
descriptions of this process are based on the assumption that the 
phase-space density $f(\vec{x},\vec{p},t)$ is to zeroth order isotropic and
independent of the pitch angle
 $\mu=\cos\alpha=\vec{p}\cdot\vec{B}/(p\,B)$ between the particle momentum
 $\vec{p}$ and magnetic field $\vec{B}$. 
Under this assumption, the process of acceleration at a plane shock wave 
moving in $x$-direction can be described using the isotropic particle density 
$n(x,p,t) = 4\pi\,p^2\,f^{(0)}(x,p,t)$, where $p=|\vec{p}|$. The transport 
equation in a plasma at velocity $u(x,t)$ is then given by 
(e.g.~Parker 1965; Jones \& Ellison 1991):
\eqb\label{diff_konv}
\frac{\partial n}{\partial t} +  \frac{\partial}{\partial x}(u\,n + F) =
   \frac{1}{3}\frac{\partial u}{\partial x}
   \frac{\partial}{\partial p}(p\,n) \,,
\eqe
where $F(x,p,t)$ is the flux due to the stochastic propagation of particles in
configuration space. 
If the spatial transport can be described by standard diffusion, then 
$F$ is proportional to the gradient in the density, and 
Eq.~(\ref{diff_konv}) is the well known diffusion-convection equation.
In this case, the momentum dependence of the phase-space density of particles 
accelerated at a strong shock is given by a power 
law $f(p)\propto p^{-s}$ with spectral index $s=3r/(r-1)\,$,
depending solely on the compression ratio\footnote{$r=\rho'/\rho$ 
with $\rho'$ and $\rho$ are the downstream and upstream plasma densities 
respectively.} $r$ of the shock.
However, the presence of a braided magnetic field (Jokipii \& Parker 1969), 
can introduce a non-diffusive spatial transport. This is important especially 
in quasi-perpendicular shock fronts, where the mean magnetic field $\vec{B}_0$ 
lies in the plane of the shock, and a stochastic component with
$\delta b:=\langle |\delta\vec{B}|\rangle/|\vec{B}_0| \ll 1$
parallel to the shock normal exists (in $x$-direction). 
Particles which follow the field lines are subject to a 
combined diffusion process. One is along the field line due to pitch-angle
scattering and the other is introduced by the stochastic spatial fluctuations
of the magnetic field on a larger scale as those responsible for scattering. 
This together leads to an anomalous transport of
particles while gaining energy due to shock crossings, which is outlined in
Sect.~\ref{transport}, followed by a brief description our Monte-Carlo 
method in Sect.~\ref{montecarlo}. This method is designed to investigate 
test-particle acceleration in magnetic fields with a stochastic component, 
without a priory 
assumptions about the pitch-angle distribution of the phase-space density. 
The results are presented in Sect.~\ref{acceleration}, showing especially
the dependence of the spectral index $s$ on the compression ratio $r$ in 
two different transport regimes in comparison to analytical treatments.

\section{Anomalous transport}
\label{transport}
The main aspect of particle transport in a braided field 
(stochastic field with $\delta b \ll 1$) is the introduction of
memory to the particle propagation. The change of the density 
at time $\tilde{t}$ is no longer proportional to the second derivative of 
the density at this time alone (standard diffusion equation), but 
also depends on the second derivative at times $t<\tilde{t}$.
This arises, because any local 
variation of the particle density which is caused by the geometry of the
magnetic field itself is {\em not} the source of a diffusive particle flux, 
and remains associated with the field line. This contribution has therefore 
to be subtracted from the standard diffusion term. 
A formulation of this kind of anomalous
transport has been given by Balescu (1995). Effectively this introduces a 
memory of the particle density at time $\tilde{t}$, on the spatial 
realisation of the magnetic field 
to which a particle was correlated during $\tilde{t}-\tcorr < t <\tilde{t}$.
An important consequence of transport in braided fields is revealed by the
time dependence of the mean quadratic deviation perpendicular to $\vec{B}_0$, 
in $x$-direction $\langle(\Delta x(t))^2\rangle \propto t^{\alpha}$, which is 
given by $\alpha=1/2$ (Rechester \& Rosenbluth 1978; Rax \& White 1992).
This kind of transport is called {\em sub}-diffusion. Note that here $x$ is the
direction along the shock normal, and the relevant reference system is the
magnetic field, which flows downstream with the background plasma. 
The time dependence shows that particles are even more effectively swept 
away from the shock in downstream direction as compared to standard 
diffusion ($\alpha=1$). 
This increases the escape probability and leads
to a steeper spectrum as shown by the results in Sect.~\ref{acceleration}.

\section{Monte-Carlo method}
\label{montecarlo}
The simulation of particle acceleration in a stochastic field
(static in the background plasma) has to consider the memory introduced by
the magnetic field as described in the previous section. The spatial
transport is a non-Markovian process. 
We generate a constant mean magnetic 
field and stochastic fluctuations at equidistant grid points, and assume the 
field to be linear in between. Using a random number generator for 
the stochastic
\begin{wrapfigure}[20]{r}{9.5cm}
\vspace{6.5cm}
\includegraphics{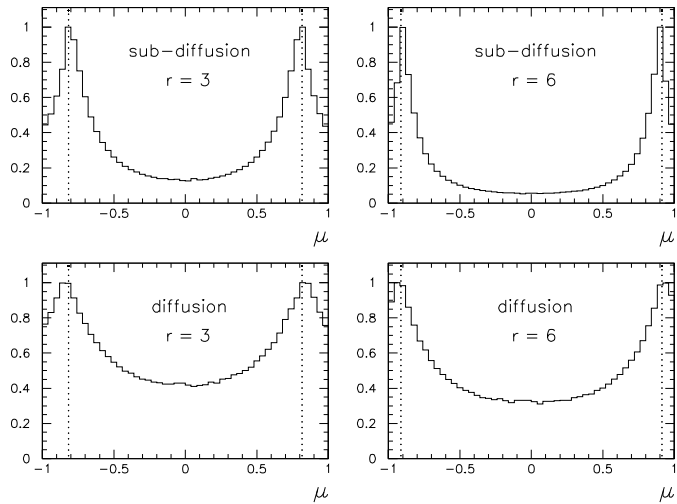}
\caption{\it Upstream pitch-angle distribution for {\em sub}-diffusion 
(upper) and diffusion (lower), for two compression ratios. The dotted lines
indicate an approximation of the maximal pitch angle $|\mu|$ for which 
reflection off the shock is possible.}\label{pitch}
\end{wrapfigure} 
 fluctuations, which allows to recall all values, 
we are able to assure a complete memory of the field until the particle 
crosses an escape boundary far downstream (Gieseler et al. 1997). 
At the same time this method allows to use a new random number for each 
field patch the particle crosses, which leads to the standard diffusion. 
A combination of the recalled 
value and a new random component would simulate a finite correlation time
of particle and magnetic field. This is, of course, the more realistic case.
However, to investigate the principal effect of {\em sub}-diffusion, 
we present here only results of 'pure' {\em sub}-diffusion and standard 
diffusion. Particles move along the field lines under the influence of 
pitch-angle scattering.
The length scale of the grid spaces of the field sampling is chosen to be 
in the same order as the scattering length. 
This assures, that while particles are transported in 
configuration space due to the field line geometry, they diffuse along the 
field line itself. At the same time this avoids, that particles diffuse
along the field while sampling only a linear patch of it.
At a change of the magnetic field direction (in particular at the 
shock) we make use of the conservation of the magnetic moment
 $\pperp^2/B$, where $\pperp$ is the component of the momentum perpendicular
to the magnetic field $B$. This approximation is valid especially for
non-relativistic quasi-perpendicular shocks, which we consider here 
(Gieseler et al. 1999, see also for a description of pitch-angle scattering).
The momentum remains constant in the corresponding upstream and 
downstream rest frames. On crossing the shock the momentum and pitch-angle
is transformed into the new system (Gieseler 1998).
This method allows to measure the particle propagator and the steady 
state density profile, which are in agreement with theoretical
predictions from Kirk et al. (1996) for {\em sub}-diffusive transport 
(Gieseler et al. 1997). Furthermore we are able to measure the pitch-angle 
distribution and the momentum spectrum which are presented in the next
section.

\section{Particle acceleration at quasi-perpendicular shocks}
\label{acceleration}
\begin{wrapfigure}[20]{r}{10cm}
\vspace{6.5cm}
\includegraphics{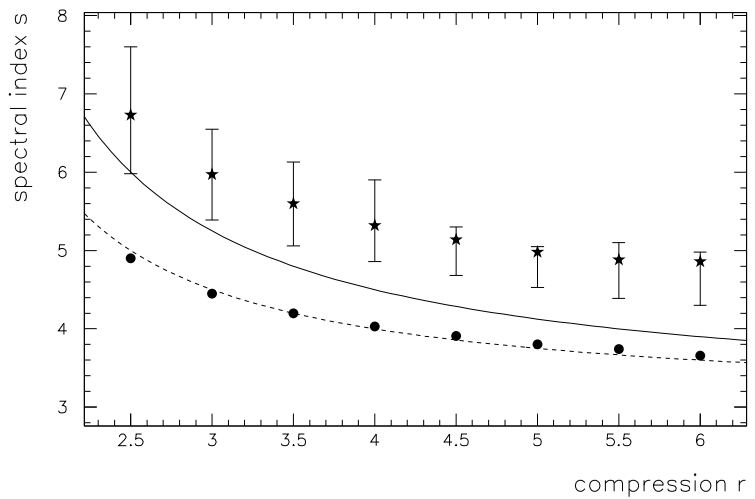}
\caption{\em Spectral index $s$ vs. compression ratio $r$. Discrete symbols
represent our Monte-Carlo results. Lines represent analytical results for
isotropic phase-space distributions 
(see text). Stars and solid line: {\em sub}-diffusion. 
Dots and dashed line: diffusion. }\label{index}
\end{wrapfigure}

In accelerating particles over between two orders of magnitude
(for the steepest spectra) and six orders of magnitude
we always find a power law for the momentum distributions. We do not include
loss mechanisms, and fit a power law function $f\propto p^{-s}$ between about
one order of magnitude
above the injection momentum and one order of magnitude below the (technical)
cut-off. The results are plotted in Fig.~\ref{index} for relativistic
particles ($v=c$) at non-relativistic shocks ($u_{\rm s}\ll c$) for 
various compression ratios $r$. Dots represent standard diffusive
acceleration, where the value of the fluctuation of a patch of field line is 
always random, i.e.~no memory effect is introduced. The stars show the 
spectral index for particles which move 
always along the same field line, so that
{\em sub}-diffusive
behaviour can take effect. The statistical error of the fit itself is well
represented by the marker symbols. However, whereas the flatter diffusive
spectra extend over many orders of magnitude, the steep {\em sub}-diffusive
spectra are more difficult to measure. The maximal systematical error
in finding the spectral index from the momentum distribution
is indicated by error bars. Because the memory effect for 
{\em sub}-diffusion can not set in immediately, the momentum distribution
has a plateau below about ten times the injection momentum. This is indicated 
by the lower bound of the error bar. A fit to the region where the spectrum
is cut off due to technical reasons gives the upper bound of the error bar.
For spectra flatter than about $s=5$, a cut-off is effectively absent,
so that the upper bound almost coincides with the plotted index.
It can be seen from Fig.~\ref{index}, that the spectrum for
{\em sub}-diffusive acceleration is significantly steeper than for
standard diffusion. We now compare our results to analytical predictions,
remembering that these are found under the assumption 
of an almost isotropic pitch-angle distribution. For standard diffusion
the result was referred in connection 
with Eq.~(\ref{diff_konv}): $s=3r/(r-1)\,$, and plotted as a dashed line
in Fig.~\ref{index}. Although we found the pitch-angle distribution 
is not really isotropic in this case, the spectral index found by the
Monte-Carlo method agrees quite well with the analytical result.
For {\em sub}-diffusive transport, an analytical solution was found by
Kirk et al. (1996): 
\eqb\label{spektralindex}
s=3\left(1+\frac{n(-\infty)}{n(0)} \frac{1}{r-1} \right)\,
=\frac{3r}{r-1}\left(1+\frac{1}{2r}\right)\,;\quad \mbox{where}\quad
\frac{n(0)}{n(-\infty)}=\frac{2}{3}\,.
\eqe
The second relation means that the density of continuously 
injected particles at the shock is less than the density far 
downstream. The resulting spectral index $s(r)$ is plotted as a solid 
line in Fig.~\ref{index}. 
Again, this result was found under the assumption of an almost 
isotropic phase-space density. The Monte-Carlo method does not make any 
assumptions
on this distribution, moreover we are able to measure the pitch-angle 
distribution at any distance from the shock. Figure~\ref{pitch} shows
the pitch-angle distribution immediately upstream of the shock, in the upstream
rest frame for the {\em sub}-diffusive and diffusive transport regime, at
compression ratio $r=3$ and $r=6$. Especially for {\em sub}-diffusive transport
and high compression ratio, we found the highest anisotropy. Here,
the deviation of the Monte-Carlo results from the analytical result 
(\ref{spektralindex}) is most prominent (see Fig.~\ref{index}). 
We found, that the density of accelerated particles is not only reduced
at the shock
by the amount predicted by Kirk et al. (1996), moreover a jump
arises at the shock, which is intimately related to an anisotropic 
phase-space distribution (Gieseler et al. 1999). 
This jump is such, that the upstream density is even more reduced, 
than indicated by Eq.~(\ref{spektralindex}) (Gieseler 1998). This leads to an 
increased escape probability, and therefore to a steeper spectrum, as compared
to the analytical result.

\section{Conclusions}
We presented Monte-Carlo simulations of particle acceleration
at non-relativistic quasi-perpendicular shock fronts. We found that a 
stochastic component in addition to the mean magnetic field introduces
{\em sub}-diffusive particle transport. The transport aspects (like 
propagator and density) were compared
to analytical treatments earlier (Gieseler et al. 1997), and we found very 
good agreement. Moreover, we tested our Monte-Carlo code for
oblique shocks against semi-analytical results, and found precise
agreement again (Gieseler et al. 1999).
Here we showed, that particle acceleration under the {\em sub}-diffusive
transport regime leads to a much steeper spectrum (e.g.~$s=5.3$ for $r=4$)
compared to standard diffusion ($s=4.0$ for $r=4$), even steeper than
predicted by Kirk et al. (1996).
The steepening of the spectrum depends strongly on whether or not
particles are correlated to field lines, and not (to first order) on the 
shock velocity, the scattering operator, or the amplitude of the magnetic 
field fluctuations.
However, if the mean field is not strictly perpendicular, i.e.~is
oblique with an angle $\Theta$ with respect to the shock normal, then the 
transport properties depend on the amplitude of the fluctuations.
The {\em sub}-diffusive transport will take effect as long as
 $\delta b> 1/\tan\Theta$. It is clear, that for the {\em sub}-diffusive
transport regime our result yields an upper limit on the spectral index
(i.e.~the steepest possible), because it was produced by an unlimited 
correlation of particle and field line. In reality, particles will, of course,
decorrelate from a given initial field geometry. This is connected with 
the realisation of the magnetic field itself, and is subject to further
investigation.

\section{Acknowledgments}
This work was supported by the University of Minnesota
Supercomputing Institute, by NSF grant AST-9619438 and 
by NASA grant NAG5-5055. U.G. acknowledges support from the 
Deutsche Forschungsgemeinschaft under SFB~328.
%
\vspace{1ex}
\begin{center}
{\Large\bf References}
\end{center}
Balescu R., 1995, Phys. Rev. E 51, 4807\\
Gieseler U.D.J., 1998, Dissertation, Univ. Heidelberg, MPI f\"ur Kernphysik, 
preprint MPI H-V6-1998\\
Gieseler U.D.J., Duffy P., Kirk J.G., Gallant Y.A., 1997, Proc. 
25. Int. Cosmic Ray Conf., Durban, 4, 437\\
Gieseler U.D.J., Kirk J.G., Gallant Y.A., Achterberg A., 1999, A\&A 345, 298\\
Jokipii J.R., Parker E.N., 1969, ApJ 155, 777; 799\\
Jones F.C., Ellison D.C., 1991, Space Science Reviews 58, 259\\
Kirk J.G., Duffy P., Gallant Y.A., 1996, A\&A 314, 1010\\
Parker E.N., 1965, Planet. Space Sci. 13, 9\\
Rax J.M., White R.B., 1992, Phys. Rev. Lett. 68, 1523\\
Rechester A.B., Rosenbluth M.N., 1978, Phys. Rev. Lett. 40, 38 
\end{document}